\pdfoutput=1 

\documentclass[utf8]{FrontiersinVancouver} 

\usepackage{url,hyperref,lineno,microtype,subcaption}
\usepackage{csquotes}
\usepackage[onehalfspacing]{setspace}
\usepackage[inkscapelatex=false]{svg}
\usepackage{booktabs}

\usepackage{comment}

\def\keyFont{\fontsize{8}{11}\helveticabold }
\def\firstAuthorLast{{da~Costa-Luis} {et~al.}} 
\def\Authors{
Casper {da~Costa-Luis}\,$^{1}$, 
Matthias J. Ehrhardt\,$^{2}$, 
Christoph Kolbitsch\,$^{3}$,
Evgueni Ovtchinnikov\,$^{1}$,
Edoardo Pasca\,$^{1}$,
Kris Thielemans\,$^{4,5,*}$,
and Charalampos Tsoumpas\,$^{6}$}

\def\Address{
$^{1}$Science \& Technology Facilities Council, UK Research \& Innovation, Rutherford Appleton Laboratory, UK \\
$^{2}$Department of Mathematical Sciences, University of Bath, UK \\
$^{3}$Physikalisch-Technische Bundesanstalt (PTB), Braunschweig and Berlin, Germany \\
$^{4}$Institute of Nuclear Medicine, University College London, London, UK
\\
$^{5}$UCL Hawkes Institute, University College London, London, UK
\\
$^{6}$Department of Nuclear Medicine and Molecular Imaging, University Medical Center Groningen, University of Groningen, Groningen, The Netherlands}
\def\corrAuthor{Kris Thielemans, Institute of Nuclear Medicine,  UCL Hospital Tower 5, 235 Euston Road, London NW1 2BU,  UK}
\def\corrEmail{k.thielemans@ucl.ac.uk}

\newcommand{\set}[1]{\mathcal{#1}}

\begin{document}
\onecolumn
\firstpage{1}

\title[PETRIC]{PET Rapid Image Reconstruction Challenge (PETRIC)} 

\author[\firstAuthorLast ]{\Authors} 
\address{} 
\correspondance{} 

\extraAuth{}

\maketitle

\begin{abstract}
\section{Introduction:} We describe the foundation of PETRIC, an image reconstruction challenge to minimise the computational runtime of related algorithms for Positron Emission Tomography (PET).

\section{Purpose:} Although several similar challenges are well-established in the field of medical imaging, there have been no prior challenges for PET image reconstruction.

\section{Methods:} Participants were provided with open-source software for implementation of their reconstruction algorithm(s). We defined the objective function and reconstruct “gold standard" reference images, and provided metrics for quantifying algorithmic performance. We also received and curated phantom datasets (acquired with different scanners, radionuclides, and phantom types), which we further split into training and evaluation datasets. The automated computational framework of the challenge is released as open-source software.

\section{Results:} Four teams with nine algorithms in total participated in the challenge. Their contributions made use of various tools from optimisation theory including preconditioning, stochastic gradients, and artificial intelligence. While most of the submitted approaches appear very similar in nature, their specific implementation lead to a range of algorithmic performance.

\section{Conclusion:} As the first challenge for PET image reconstruction, PETRIC's solid foundations allow researchers to reuse its framework for evaluating new and existing image reconstruction methods on new or existing datasets. Variant versions of the challenge have and will continue to be launched in the future.

\tiny
 \keyFont{ \section{Keywords:} image reconstruction, medical imaging, open source, challenge, inverse problems} 
\end{abstract}

\section{Introduction}
Positron emission tomography (PET) is an advanced imaging technique which uses photon pairs (produced by positron annihilation) to detect the approximate location of the positron emissions. Even with the current scanners' fast PET detectors and electronics, the measured photon pairs still pose an inverse problem \cite{SchrammFNM2022}. For this purpose, several tomographic image reconstruction algorithms have been developed over the years, first by Hounsfield and Cormack, joint recipients of the 1979 Nobel Prize for Physiology or Medicine. There are three (not necessarily distinct) categories of reconstruction algorithms \cite{gong2021EvolutionImageReconstruction}: analytical \cite{Radon,townsend1993ImagereconstructionMethodsPositron}, iterative (or statistical) \cite{qi2006IterativeReconstructionTechniques}, and artificial intelligence (AI) \cite{reader2020DeepLearningPET}. Iterative algorithms are currently preferred in clinical PET systems due to their detailed modelling of the acquisition processes (though AI-based algorithms are gradually being made clinically available). However, iterative algorithms can be computationally demanding. This is exacerbated by recent developments such as long axial field of view PET scanners, which provide much more data. Furthermore, some PET applications (such as dynamic imaging or motion correction) require subdividing acquisition data \cite{Chungjnumed.124.268706}. This results in multiple low-count reconstructions. The problem setting is therefore becoming increasingly challenging. In real practice, computational speedup in image reconstruction can permit running more complex (previously prohibitively slow) algorithms to provide significantly improved medical images.

Several advanced image reconstruction methods have been developed exploiting improved optimisation methods, machine learning, and more powerful computer hardware. However, it is difficult to objectively compare and assess such reconstruction algorithms and frameworks for accuracy and computational time.

Over the last two decades, several scientific communities have established the concept of “challenges" to facilitate fair comparison of different algorithms. These are well-known in the fields of image registration, segmentation, and reconstruction. Each challenge can focus on a particular topic in order to accelerate research in a specific direction. Inspired by these developments, in 2024 we held the PET Rapid Image Reconstruction Challenge: \href{https://www.ccpsynerbi.ac.uk/petric}{PETRIC}. In this paper, we describe the design and rationale of the challenge.

\section{PETRIC description}
The primary aim of PETRIC was to stimulate research \& development of faster PET image reconstruction algorithms applicable to real-world data. The challenge used a predetermined maximum a-posteriori (MAP, also known as penalised likelihood) problem and a “gold-standard" reconstruction algorithm which is known to converge to the optimum. Participants were given access to a sizeable set of phantom data acquired on a range of clinical scanners. The main task was to reach a solution close to the “gold standard" images (similar -- within a threshold -- mean radioactivity values within specified regions of interest) as quickly as possible (minimum computation time). One key component of the challenge was the use of the \href{https://www.ccpsynerbi.ac.uk/softwareframework-html}{Synergistic Image Reconstruction Framework (SIRF)} as a common computational platform for all participating reconstruction algorithms \cite{Ovtchinnikov2020a}. Additionally, the \href{https://ccpi.ac.uk/cil/}{Core Imaging Library (CIL)} was used to subclass its iterative algorithm class \cite{Jorgensen21}. All algorithms therefore use the same projectors and objectives, and thus their performance is determined purely by algorithmic approach.

Seven teams registered for the challenge, with four teams submitting their work for evaluation. These four teams submitted a total of nine algorithms (each team contributing up to three entries).

The design, preparation and execution of PETRIC involved several coordinated major tasks: 
\begin{itemize}
    \item Gold standard definition
    \item Collection \& curation of datasets
    \item Objective evaluation
    \item Definition of metrics \& thresholds
    \item Live scoring \& feedback
    \item Concluding workshop \& awards
\end{itemize}

\section{Materials and methods}

\subsection{Reconstruction problem}
\label{sec:optimisation-problem}

The optimisation problem is a MAP estimate using the smoothed relative difference prior (RDP) \cite{nuyts2002concave}, i.e.
\begin{align}
    x^* = \arg\max_{x \in \set C} \Psi(x; y) 
\end{align}
with $y$ the measured data, $x$ the reconstruction image and
\begin{align}
\Psi(x; y)=L(y; \hat y(x)) - \beta R(x)
\label{eq:objective_function}
\end{align}
where the constraint set $\set C$ is defined as 
\begin{align}
x \in \set C \quad \Leftrightarrow \quad \begin{cases} x_i \geq 0 & i \in \set M \\ x_i = 0 & \text{else}
\end{cases}
\end{align}
for some mask $\set M$ provided. The parameter $\beta > 0$ is the regularisation parameter that trades-off the fit to the data with the regularity provided by the prior. Note that SIRF defines the objective function as above, but algorithms from CIL minimise the function. The log-likelihood (up to terms independent of the image) is
\begin{align}
L(y; \hat y) = \sum_k y_k \log \hat y_k - \hat y_k
\end{align}
with $y$ a vector with the acquired data (histogrammed), and $\hat y$ the estimated data for a given image $x$,
\begin{align}
\hat y(x) = \operatorname{diag}(m)(Ax + a).
\label{eq:forward_model}
\end{align}
Here $m$ denote the multiplicative factors corresponding to the detection efficiencies and the attenuation coefficient factors, $a$ the “additive” background term corresponding to an estimate of randoms and scatter coincidences, pre-corrected with $m$), $A$ is an approximation of the line integral operator \cite{schramm2024parallelproj}, and $\operatorname{diag}$ an operator converting a vector to a diagonal matrix.

Due to PET conventions, for some scanners, some data bins will always be zero, which corresponds to any potential “virtual crystals”, in which case any corresponding elements in $m$ will also be zero. 
The corresponding term in the log-likelihood is defined as zero. 
All other elements in $a$ are guaranteed to be (strictly) positive $(a_i > 0)$.

The smoothed RDP is given by
\begin{align}
    R(x) = \frac12 \sum_{i \in \set V} \sum_{j \in \set N_j} w_{i,j} \kappa_i \kappa_j \frac{(x_i - x_j)^2}{x_i + x_j + \gamma |x_i - x_j| + \varepsilon},
    \label{eq:RDP}
\end{align}
with $\set V$ the set of voxels, $\set N_i$ the neighbourhood of voxel $i$ (here taken as the 26 nearest neighbours), $w_{i,j}$ weight factors (here taken as “horizontal" voxel-size divided by Euclidean distance between the $i$ and $j$ voxels), $\kappa_i$ an image to give voxel-dependent weights (here predetermined as the square root of minus the row-sum of the Hessian of the log-likelihood at an initial ordered subsets expectation maximisation (OSEM) \cite{hudson1994accelerated} reconstruction, see \cite[Equation (25)]{tsai2019benefits}, $\gamma$ an edge-preservation parameter (here taken as 2), and $\varepsilon$ a small number to ensure smoothness (here predetermined from an initial OSEM reconstruction). The initial OSEM reconstruction was also given to all participants to potentially use as a starting point.

\subsection{Reference reconstruction}
\label{sec:reference-reconstruction}

\begin{figure}[htbp]
\centering
\includegraphics[width=0.3\textwidth]{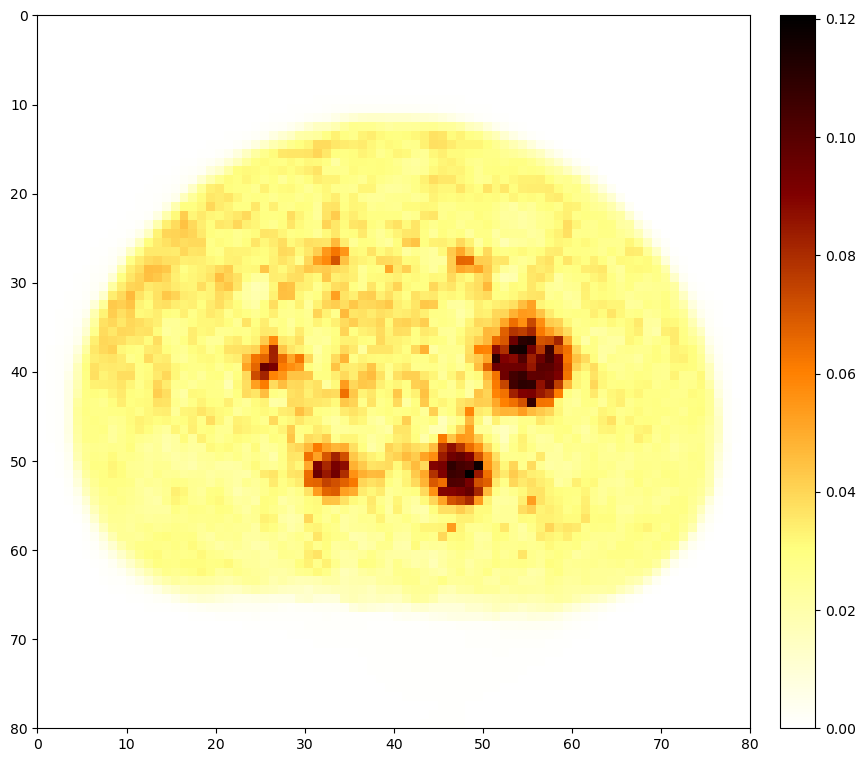} \hspace{5mm}
\includegraphics[width=0.3\textwidth]{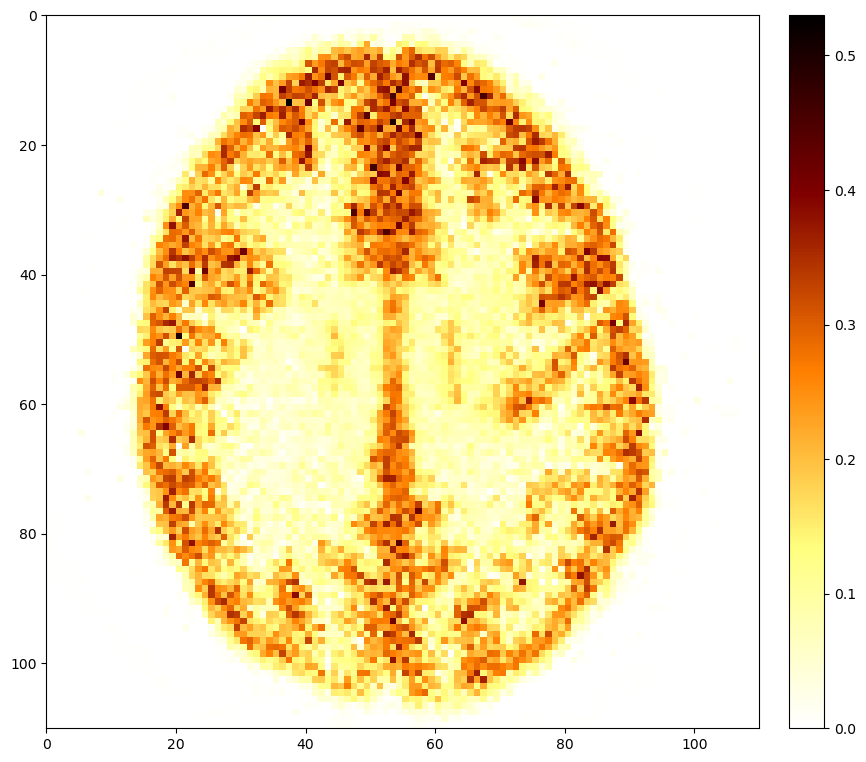}

\caption{
Reference images for two phantoms: on the left the low counts NEMA scanned with the Mediso clinical PET scanner, on the right the Hoffman phantom scanned with a Siemens Biograph Vision 600. The images shown are cropped to the location of the phantom.
} \label{fig:reference:osem}
\end{figure}

Reference MAP solutions are computed using a modified version of the BSREM algorithm, using an EM-like preconditioner, subsets, and a relaxed step-size:
\begin{align}
\hat x^\mathrm{new} =  \set P_{\set C}  \left(x + \alpha \frac{(x + \delta)}{S} \operatorname{\nabla}_x^s \Psi(x; y)\right),
\end{align}
with $\set P_{\set C}$ the projection onto the non-negative image space and $S$ the average of the “sensitivity images" $A_s^T \cdot m_s$ for each subset $s$, $\delta$ a small number (determined from the OSEM image) and $\alpha$ a step-size that decreased with iterations. This algorithm has been shown to converge to the MAP solution \cite{ahn2003globally}.  Our implementation \cite{SIRF_Contribs_BSREM} used a number of subsets different for  each specific scanner, and manual tuning of the step size parameters.
Nevertheless, we found that a high number of updates (of the order of 15\,000) was often required. If necessary, the algorithm was restarted until our convergence metrics were satisfied, see section \ref{sec:metrics}. Two reference reconstructions for test datasets are shown in Figure~\ref{fig:reference:osem}. A dedicated example demonstrating PET image reconstruction with BSREM using SIRF can be found in a corresponding SyneRBI \href{https://github.com/SyneRBI/SIRF-Contribs/blob/master/src/notebooks/BSREM_illustration.ipynb}{notebook}.

\subsection{PETRIC Datasets}

\subsubsection{Data Collection}
At the start of the challenge, we provided example acquired data of a small set of phantoms for participants to use for the development of their methods, complemented by further datasets during the challenge. Example test datasets are shown in Figure~\ref{fig:reference:osem}. An overview of the curated data is in Table \ref{tab:training_data}.  New datasets were acquired at different sites following the challenge training period, see Table \ref{tab:evaluation_data}. These datasets were used for scoring the performance of each algorithm. In this way the bias towards a certain vendor or scanner model was kept minimal. In addition, this approach ensured fairness of the competition because at the time of the challenge, nobody had access to the final ground truth data - including the organisers themselves. 

We requested datasets from several medical physics groups. All datasets had to be made open access, under the Creative Commons Attribution 4.0 International license.
We encouraged the contributing scanning sites to provide not only the acquired raw data for the final testing but also the corresponding regions of interest. All phantom datasets have been made publicly available after the formal completion of the challenge. Participants were free to test their algorithms on additional datasets, for instance those available at the Zenodo SyneRBI Community. Human datasets were not included due to difficulties with organising data sharing agreements and privacy issues related to potential use of the data for facial recognition~\cite{Selfridge1304}.

Only simple data from scanners and acquisition modes (e.g., only one bed position) were considered in this challenge. Reading the data was supported by the open source software for tomographic reconstruction \cite[\href{http://stir.sf.net}{STIR}]{Thielemans2012P} either directly or through one of our conversion scripts. Some of these scripts have not been made publicly available yet due to pending approval from scanner vendors. The scanner models for which we can handle data are:
\begin{itemize}
    \item Siemens Healthineers: Biograph mMR (native SIRF/STIR processing), Biograph Vision 600 (conversion via e7tools)
    \item GE HealthCare: Signa PET/MR (conversion via Duetto \cite{wadhwa2018ImplementationImageReconstruction}, but we have no access to the MRAC code), Discovery 690/710 (conversion via pettoolbox), Discovery MI (conversion via Duetto)
    \item Mediso: AnyScan (conversion handled by the National Physical Laboratory (NPL))
    \item Positrigo: NeuroLF from output in SAFIR file format \cite{dao2022EvaluationSTIRLibrary} 
\end{itemize}

All necessary datasets were saved in a STIR-supported version of the Interfile standard to allow offline reconstruction. A readme file with information about the phantom preparation and acquisition protocols was provided. 

The data collected included the following files: 
\begin{itemize}
    \item list-mode files, as exported from the scanner console,
    \item attenuation correction files as provided by the scanner (i.e., CTAC or MRAC) in DICOM, Nifti or STIR Interfile format,
    \item normalisation and calibration files, in the format supplied by the vendor,
    \item image reconstructed with vendor software (DICOM) with all corrections applied for cross-checking,
    \item volumes of interest (VOI) provided as images in exactly the same dimensions / voxel-size as the scanner reconstruction (Nifti or STIR Interfile format), one file for each VOI. VOIs were constructed by the sites contributing the data or constructed by the PETRIC team. Scripts were provided to construct VOIs for the NEMA Image Quality and Hoffman phantoms.
\end{itemize}
Curated datasets are available at \href{https://petric.tomography.stfc.ac.uk/data/}{petric.tomography.stfc.ac.uk/data/} 
\begin{table}
    \centering
    \begin{tabular}{lllcc}
        \toprule
        scanner model & phantom  & radionuclide & low count &  reference \\
        \midrule
       Siemens mMR  & NEMA IQ  & F-18 & & \cite{thomas_2018_1304454}\\
       Siemens mMR  & NEMA IQ  & F-18 & \checkmark & \cite{ucl_2025_16920471}\\
       Siemens mMR  & ACR      & F-18 &  & \cite{pawel_j_markiewicz_2021_5760092}\\
       Positrigo NeuroLF  & Hoffman  & F-18 & & \cite{university_hospital_leipzig_2025_16920235,university_hospital_leipzig_2026_18415418}\\
       Siemens Vision 600 & Alderson thorax  & F-18 & & \cite{leek_2026_18434368}\\
       Mediso Anyscan & NEMA IQ & Ge-68 & & \cite{deidda_2024_13946493}\\
       GE Discovery MI 3-ring & Antropomorphic torso  & F-18  & & \cite{GE_DMI3_Torso}\\
       \bottomrule
    \end{tabular}
    \caption{Training data made available to participants during the challenge.}
    \label{tab:training_data}
\end{table}
\begin{table}
    \centering
    \begin{tabular}{lllcc}
        \toprule
        scanner model & phantom & radionuclide  & low count &  reference \\
        \midrule
       Positrigo NeuroLF  & Esser & F-18   & & \cite{university_hospital_of_zurich_2026_18412718}\\
       Siemens Vision 600 & Hoffman & F-18 & & \cite{li_2026_18374965}\\
       Siemens Vision 600 & NEMA IQ & Zr-89 & \checkmark & \cite{li_2026_18380638}\\
       Mediso Anyscan & NEMA IQ & Ge-68 & \checkmark & \cite{deidda_2025_16919133}\\
       GE Discovery 690 &  NEMA IQ & F-18 & & \cite{Turku_IQ_Jan_2026_D690}\\
       GE Discovery MI 4-ring & NEMA IQ & F-18 & & \cite{Turku_IQ_Jan_2026_DMI}\\
       \bottomrule
    \end{tabular}
    \caption{Evaluation data released after the submission deadline.}
    \label{tab:evaluation_data}
\end{table}

\subsubsection{Data Curation}
To prepare the PET datasets for analysis, we followed a structured data curation workflow. The process was designed to ensure consistency, reproducibility, and compatibility with the SIRF framework. We describe this workflow here. More detail is provided in Appendix A.

All datasets were organised under a standardised directory structure, where the dataset name followed the convention \begin{tt}scannername\_phantomname\end{tt}. 
Initial quality control checks were performed using a dedicated script to validate the integrity of the raw data. Subsequently, we generated initial images using one of the provided VOIs as a template. Since all VOIs shared the same geometry, any of them could be used interchangeably. In cases where VOIs were not available, we used either a vendor-supplied image or a placeholder.
During image generation, some metadata—such as modality, radionuclide, and scan duration—was lost. We manually restored this information by editing the relevant image header files, referencing the original prompt data. For internal reference reconstructions, we also configured the number of subsets in the dataset settings file.
To support automated figure generation, we specified the slices of interest in the PETRIC configuration files. The center-of-mass coordinates of the VOIs, obtained during quality control, were used to guide this selection. When VOIs were not pre-defined, we generated them ourselves, based on the OSEM images.
After VOI creation, we re-ran the quality control process to produce additional plots and verify VOI alignment. To calibrate reconstruction parameters, we estimated the penalisation factor by comparing each dataset with a reference dataset acquired on the same scanner, followed by some manual tuning. 
For reference solutions, we performed BSREM reconstructions as mentioned above. Given the computational demands of this step, we monitored the reconstruction progress and image quality throughout. Once the reconstructions are completed, we adapted a plotting script to visualise BSREM metrics and copied the final reconstructed image to the designated reference location within the dataset directory.
To finalise the dataset, we removed intermediate files and verified the presence and accuracy of documentation files such as README.md. The curated datasets were then transferred to a web server for distribution and further use. 

\subsection{Evaluation and scoring}
\label{sec:evaluation_and_scoring}

We  defined several metrics which were used to determine the fastest algorithm to reach target image quality, see next section. All algorithms were executed for each dataset until their solution reached the required thresholds on all selected metrics (for $10$ subsequent updates) or their runtime exceeded one hour.
For each metric, results from all teams were ranked according to the wall-clock time required to reach the threshold on our standard computational platform. Ranking is from worst (rank~1) to best (rank~N), with best corresponding to “the fastest to reach threshold". The overall rank for each algorithm was the average of all ranks for the individual metrics on each dataset. The algorithm with the highest overall rank won the challenge. Note that due to potential variability in the wall-clock timing as well as in using the stochastic algorithms, each reconstruction was executed three times, and the median of the wall-clock time was used for the ranking.

\subsection{Metrics and thresholds}
\label{sec:metrics}


\newcommand{\RMSE}{\operatorname{RMSE}}
\newcommand{\MEAN}{\operatorname{MEAN}}

Each entry to the challenge was evaluated on metrics based on the object as a whole, the background and specific volumes of interest (VOI). They all use the voxel-wise root mean-squared-error (RMSE) of image $x$ computed in region $\set R$ with respect to the reference image $r$: $$\RMSE(x; r; \set R) = \sqrt{\frac{1}{|\set R|}\sum_{i \in \set R} (x_i - r_i)^2},$$ 
and the mean of image $x$ for region $\set R$, $$\MEAN(x; \set R) = \frac{1}{|\set R|}\sum_{i \in \set R} x_i,$$
as building blocks.

We required three conditions for an image $\theta$ to satisfy the convergence criteria. First, the RMSE of the whole object satisfies $$\frac{\RMSE(\theta; r; \set W)}{\MEAN(r; \set B)} < 0.01. $$
Second,
the background RMSE satisfies  $$\frac{\RMSE(\theta; r; \set B)}{\MEAN(r; \set B)} < 0.01. $$
Third, in each VOI $\set R_i$ the AEM (absolute error of the mean) satisfies $$\frac{|\MEAN(\theta; \set R_i) - \MEAN(r; \set R_i)|}{\MEAN(r; \set B)} < 0.005. $$
Here $r$ denotes the converged BSREM reference image, $\set W$ the marginally eroded whole object VOI, $\set B$ the background VOI and $\set R_i$ specific VOIs (“tumors”, “spheres”, “white/gray matter”, etc.).
Note that metrics therefore only took into account regions of interest in the phantom and excluded the non-radioactive background. 
The thresholds were chosen to be well-below clinically accepted variation while avoiding being too stringent and depend on numerical errors. In particular, ``converged" images were visually very similar, except possibly in the non-radioactive background.

\subsection{Infrastructure} 

To run a public competition, several components are needed: registration forms, template code, data hosting, submission method, standardised and secure GPU-enabled continuous integration (CI), and live leader boards.

We used a dedicated \href{https://github.com/SyneRBI/PETRIC}{GitHub repository} for providing a starter code, and to create private repository forks per participating team. For team registration, \href{https://github.com/SyneRBI/PETRIC/issues/new?template=participate.yml}{GitHub issue forms} were used. Hosting everything under one GitHub organisation also allowed secure access to our \href{https://docs.github.com/en/actions/concepts/runners/self-hosted-runners}{self-hosted Continuous Integration (CI) runner}, which has the required docker image \cite{PETRIC_docker,SIRF_SB_380}, training data, and a GPU. Code uploaded by teams is added to a sequential queue to run on this CI.

All of the reconstructions were run on our standard platform: 32 core CPU, 100 Gb RAM, NVIDIA A100 40Gb PCIe.  The CI machine also runs a \href{https://petric.tomography.stfc.ac.uk}{web server} from which teams could see live \href{https://petric.tomography.stfc.ac.uk/tensorboard/?smoothing=0}{TensorBoard logs} of their CI runs, and \href{https://petric.tomography.stfc.ac.uk/data/}{download training data} locally. After the end of the submission period, test datasets were added to the machine and all submissions were rerun to produce a \href{https://petric.tomography.stfc.ac.uk/leaderboard}{final leader board}.

All code for this backend machine is available in the corresponding backend repository \cite{PETRIC_backend}.

\section{Results}


\subsection{Submitted algorithms} 
\label{sec:submitted_algorithms}
Four teams submitted a total of nine algorithms for evaluation, summarised in Table \ref{tab:teams}. Their submissions are linked in the table which should be referred to for all details.
\begin{table}[htbp]
\begin{tabular}{ccp{11.5cm}}
    \hline\hline \\
    Team & Submission & High-level Summary \\
    \midrule

MaGeZ & \href{https://github.com/SyneRBI/PETRIC-MaGeZ/tree/refs/tags/ALG1}{ALG1} & SVRG with prior preconditioning \\

 & \href{https://github.com/SyneRBI/PETRIC-MaGeZ/tree/refs/tags/ALG2}{ALG2} & like ALG1 but with Barzilai--Borwein step-size rule  \\

 & \href{https://github.com/SyneRBI/PETRIC-MaGeZ/tree/refs/tags/ALG3}{ALG3} & like ALG2 but with fine-tuned deterministic subset selection \\

    \midrule
    
SOS & \href{https://github.com/SyneRBI/PETRIC-SOS/tree/refs/tags/SAGA_final2}{SAGA\_final2} & warm-started SAGA with preconditioning and Armijo line search \\

 & \href{https://github.com/SyneRBI/PETRIC-SOS/tree/refs/tags/SVRG_final}{SVRG\_final} & like SAGA\_final2 but with SVRG \\


    \midrule

Tomo-Unimib & \href{https://github.com/SyneRBI/PETRIC-Tomo-Unimib/tree/refs/tags/LP_final}{LP\_final} & diagonally preconditioned conjugated gradient descent \\

    \midrule
    
UCL-EWS & \href{https://github.com/SyneRBI/PETRIC-UCL-EWS/tree/refs/tags/EWS_GD}{EWS\_GD} & DL-warm-started Gradient descent with Barzilai--Borwein step-size and tailored EM-preconditioner \\

 & \href{https://github.com/SyneRBI/PETRIC-UCL-EWS/tree/refs/tags/EWS_SGD}{EWS\_SGD} & like EWS\_GD but with SGD and Distance over Weighted Gradients step-size  \\

 & \href{https://github.com/SyneRBI/PETRIC-UCL-EWS/tree/refs/tags/EWS_SAGA}{EWS\_SAGA} & like EWS\_SGD but with SAGA, Katyusha acceleration and DoWG step-size \\
    \hline\hline
    
\end{tabular}
\caption{Overview of submissions sorted by team name in alphabetical order.} \label{tab:teams}
\end{table}
All submissions are gradient-based iterative algorithms but differ in a range of aspects. For the core algorithm, teams used Gradient Descent (GD) with stochastic approximation (Stochastic Gradient Descent; SGD), variance reduction (SAGA, SVRG), as first adapted to PET in \cite{twyman2023InvestigationStochasticVariance}, and acceleration (Katyusha). See \cite{ehrhardt2025guide} for an overview on these concepts. All teams experimented also with preconditioning ranging from preconditioners motivated from MLEM, diagonal approximations of the RDP Hessian and various combinations of the two. Regarding step-sizes, the teams used a-priori defined decaying step-sizes as are common in subset-based algorithms like BSREM but also experimented with others like Barzilai--Borwein, Armijo line-search and the Distance over Weighted Gradients (DoWG) method \cite{khaled2023dowg}. 
Various teams considered advanced warm-starting options computed by other iterative algorithms as well as deep learning (DL) methods.

The teams also fine-tuned subset-selection methods like the number of subsets and in which order they should be picked. 
All teams agreed that the number of subsets should be scanner specific and depend on various factors like trading off computational resource and prime factors of the number of views. 
Typical chosen values are in the range of 20 to 50. 
For the actual selection, teams used randomly selected sequences with and without replacement, established deterministic sampling (e.g. Hermann--Meyer) as well as novel variants. 
It should be noted that some teams also decided to seed the randomness making their stochastic gradient descent algorithm in practice deterministic.

More details on the MaGeZ submission can be found in \cite{MaGeZ}. The Tomo-Unimib was refined after the challenge ended in \cite{colombo2025GeneralizablePreconditioningStrategies}.








\begin{figure}[htbp]
    \centering
    \includegraphics[width=0.7\linewidth]{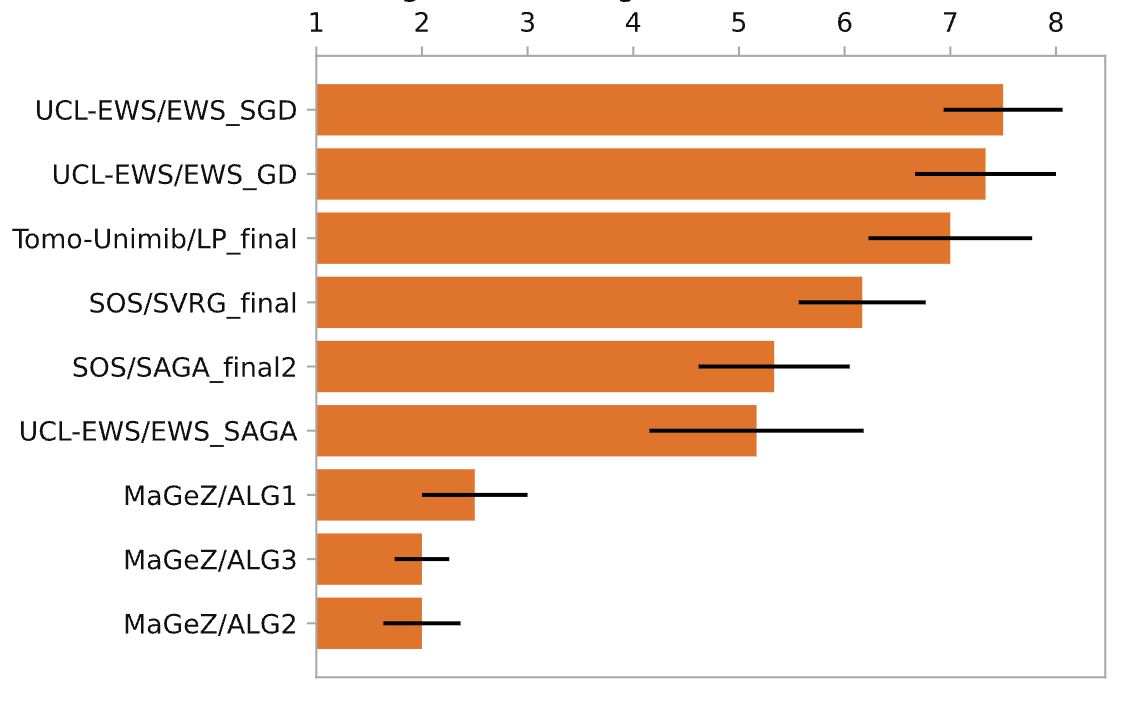}
    \caption{Mean rank of the algorithms across all datasets. The black bars show the standard error in the mean.
    \label{fig:ranks}}
\end{figure}


\begin{figure}[htbp]
\centering
\includegraphics[width=0.8\textwidth]{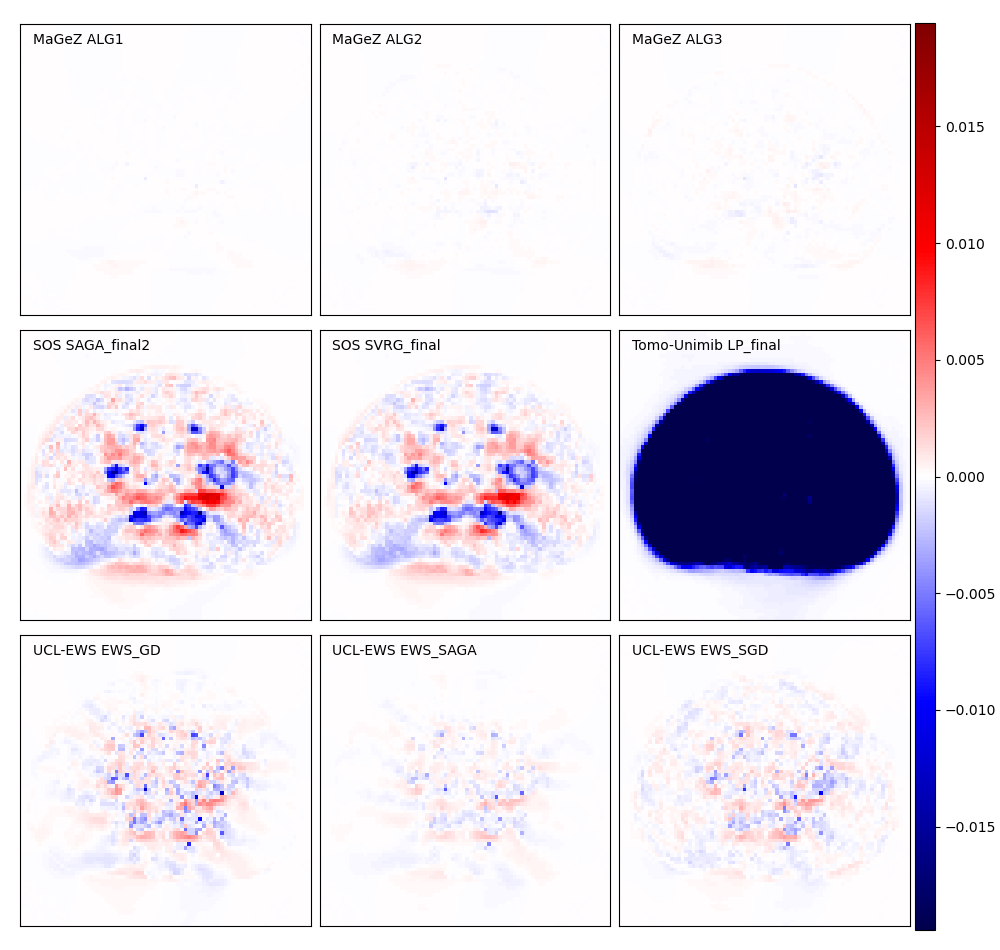}
\caption{Difference of solution obtained by all algorithms stopped when the winning algorithm reached the threshold and reference image, for the NEMA phantom low counts acquired with a Mediso scanner. 
The plots are presented in a diverging color map with white indicating a match, blue indicating underestimation and red overestimation with respect to the reference solution. 
The scale of the color map is 10\% of the maximum value of the reference image and is the same for all plots. 
The TomoUnimib algorithm did only 3 iterations and it was diverging with a result that was still the zero image. The images shown are cropped to the location of the phantom.
\label{fig:NEMA_diff}}
\end{figure}
\begin{figure}[htbp]
\centering
\includegraphics[width=0.8\textwidth]{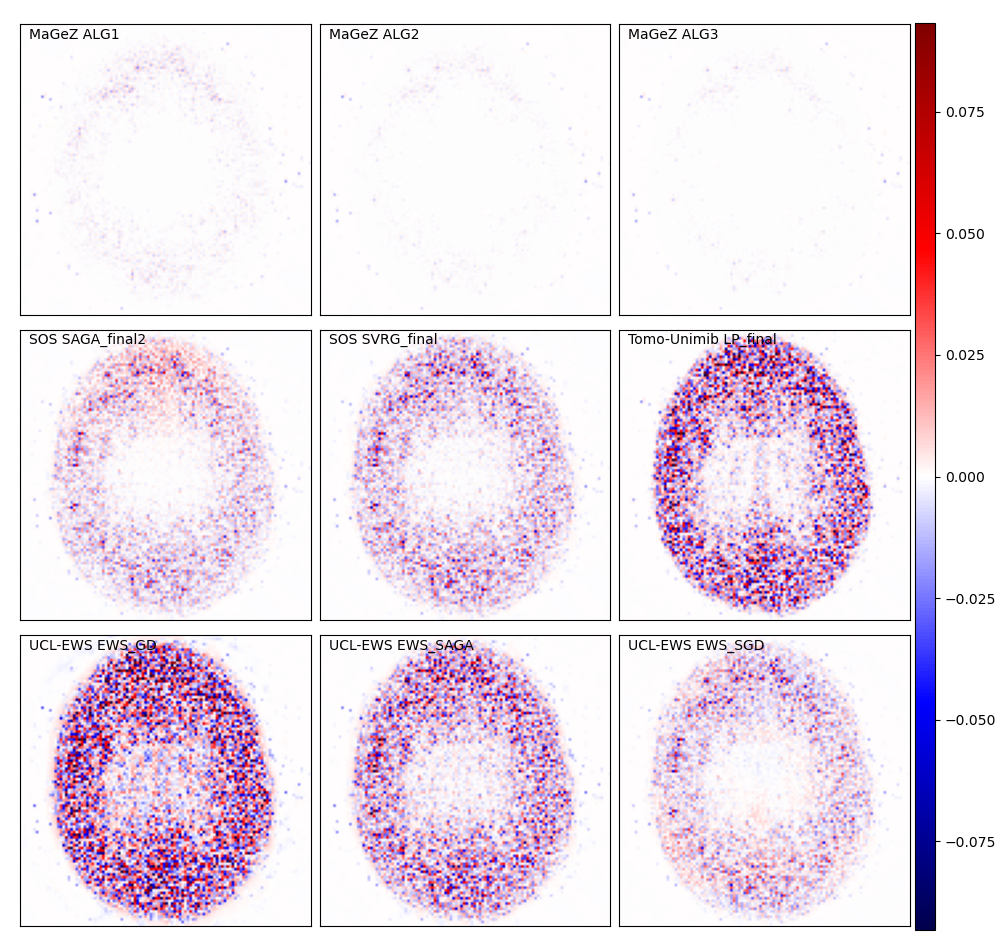}
\caption{Difference of solution obtained by all algorithms stopped when the winning algorithm reached the threshold and reference image, for the Hoffman phantom scanned with the Siemens Biograph Vision600. 
The plots are presented in a diverging color map with white indicating a match, blue indicating underestimation and red overestimation with respect to the reference solution. 
The scale of the color map is 10\% of the maximum value of the reference image and it is the same for all plots.
Here all algorithms except MaGeZ's produced a smoother image than the reference, as indicated by the seemingly random noise visible in rows 2 and 3; however, there is some structural difference in the difference plots, especially between the low and high intensity regions. 
The images shown are cropped to the location of the phantom.
\label{fig:Hoffman_diff}}
\end{figure}

\subsection{Algorithms performance and insights}
%
%

%
%
%
All submitted algorithms, see Table~\ref{tab:teams},  were run until their solution obtained a specific relative error on all selected metrics or their runtime exceeded 1 hour, see section~\ref{sec:metrics}. Details on ranking method are given in section \ref{sec:evaluation_and_scoring}.
In Figure~\ref{fig:ranks} we show the rankings of the algorithms with an error bar indicating variability in performance among runs.

The wide variety of the data provided led to varied algorithm performance across different phantom regions. 
Some algorithms performed well on some datasets, while struggling with others due to poor parameterisation, as can be seen in Figures~\ref{fig:NEMA_diff} and \ref{fig:Hoffman_diff}.
Notably, while team rankings remained consistent across all runs, there was significant variability in ranking positions among individual submissions from the same team. 
This highlights the difficulty to maintain performance across such heterogeneous datasets, with potential significative gains to be obtained by accurate parametrisation for each scanner. 
In general the algorithms from the winning team outperformed all the other algorithms.

As discussed in Section~\ref{sec:submitted_algorithms} all submitted algorithms are variants of gradient descent with stochastic approximation, which have acquired popularity recently in literature, \cite{ehrhardt2025guide}. For example, MaGeZ, UCL-EWS and SOS teams all used the SVRG algorithm, nevertheless achieving different performance due to choices in pre-conditioning, subset selections and step sizes. 

\begin{subfigure}
    \centering
    \begin{minipage}[b]{0.3\textwidth}
    \includesvg[width=\linewidth, height=4cm]{./figures/DMI4_NEMA-RMSE_whole_object.svg}
    \caption{DMI4 (whole)}
    \end{minipage}  
    \begin{minipage}[b]{0.3\textwidth}
    \includesvg[width=\linewidth, height=4cm]{./figures/NeuroLF_Esser-RMSE_whole_object.svg}
    \caption{Esser (whole)}
    \end{minipage}
    \begin{minipage}[b]{0.3\textwidth}
    \includesvg[width=\linewidth, height=4cm]{./figures/Vision600_ZrNEMA-RMSE_whole_object.svg}
    \caption{ZrNEMA (whole)}
    \end{minipage}

    \centering
    \begin{minipage}[b]{0.3\textwidth}
    \includesvg[width=\linewidth, height=4cm]{./figures/DMI4_NEMA-AEM_VOI_sphere2.svg}
    \caption{DMI4 (VoI)}
    \end{minipage}  
    \begin{minipage}[b]{0.3\textwidth}
    \includesvg[width=\linewidth, height=4cm]{./figures/NeuroLF_Esser-AEM_VOI_cold_cylinder.svg}
    \caption{Esser (VoI)}
    \end{minipage}
    \begin{minipage}[b]{0.3\textwidth}
    \includesvg[width=\linewidth, height=4cm]{./figures/Vision600_ZrNEMA-AEM_VOI_sphere2.svg}
    \caption{ZrNEMA (VoI)}
    \end{minipage}

\setcounter{figure}{5}
\setcounter{subfigure}{-1}
    \caption{Whole-object and VOI-specific normalised RMSE against time for some datasets \cite{Turku_IQ_Jan_2026_DMI,university_hospital_of_zurich_2026_18412718,li_2026_18380638}.}
    \label{fig: subfigures}
\end{subfigure}

\subsection{Outcomes of the Challenge}

The outcomes of the challenge were multifaceted. 
Beyond the competition itself, the event served as a valuable learning experience. 
One of the key takeaways was the availability of open-source implementations. 
All three top-scoring teams made their code publicly accessible. 
While some of these implementations may lack thorough documentation, they are available for exploration and reuse. This openness is a promising step toward future research and development, and some of their contributions are part of the same research topic of Frontiers in Nuclear Medicine.
For those looking to get started in this area, these implementations offer a solid foundation. 
To support reproducibility and further experimentation, we have provided Docker images \cite{PETRIC_docker} containing all necessary software.

Another key outcome is the curated phantom data collection which is available on our website, covering a variety of scanners, phantoms and tracers. Additionally, the  framework that we developed for the challenge can serve as a starting point for others interested in organising similar challenges.

During the organisation of the challenge, we also encountered and addressed several issues in our own software. This led to significant improvements, including the development of numerous scripts for data ingestion, validation, and visualisation, which may prove useful in many other instances. Major weaknesses in the performance of some  algebraic operations, as well as exposure of the underlying data arrays to Python, were identified and resolved after the challenge.

The PETRIC challenge was concluded with a corresponding focused workshop as part of the 2024 IEEE Nuclear Science Symposium and Medical Imaging Conference in Tampa, Florida, USA\footnote{Recordings of the presentations at the workshop, including of the three top-ranked teams, are available at \href{https://www.ccpsynerbi.ac.uk/petric-workshop-2024-recordings/}{www.ccpsynerbi.ac.uk/petric-workshop-2024-recordings/}}.

\section{Discussion and Outlook} 
The PETRIC challenge was successful and achieved several of its aims. The relatively low turnout of participants may be attributed to factors such as the tight timeline from the announcement of the challenge (June 10th 2024) to its start and conclusion in the same year, June 28th and September 30th respectively; but also to the partial availability of training data at the start of the challenge. 

One of the aims of this challenge was to evaluate the performance of the reconstruction algorithms on real data. The datasets in this challenge were acquired from several different scanner models and vendors, and different phantoms to cover a wide range of real-world scenarios. The data varied strongly in terms of imaging parameters (e.g., time-of-flight versus non-time-of-flight) and noise (e.g., due to different scan durations and phantom fillings). This posed a significant challenge for all the submitted algorithms because they had to perform well for a very heterogeneous set of data.

This allowed us to assess how well the performance of the different algorithms generalises to different applications. As mentioned above, to ensure the challenge was as fair as possible, data used for the final scoring was obtained after the submission deadline of the challenge. This was only possible due to the generous contribution of data from several different research sites.

One major difficulty with using data from different scanners was to convert the vendor data to a common open data format which is compatible with SIRF and STIR, respectively. In this occasion we used STIR interfile format for PET projection data. Some vendors and scanner models required closed-source tools only available based on research agreements with the vendors and detailed knowledge about the vendor data formats. This restricted the number of research sites which could contribute datasets to the challenge.
 In the future more scanners may become available thanks to the \href{https://etsinitiative.org/}{new initiative} on emission tomography standardisation, of which one aim is to design a new file format, the PET Standardisation Initiative’s Raw Data (PETSIRD) \cite{thielemans2025FinalDraftPETSIRD}.

Using real data meant that no ground truth reference solution was available and therefore the scoring of algorithms needed careful design. Rather than setting up a challenge which looks for the algorithm yielding the best possible image, we designed the challenge to find the algorithm which reaches a known solution in the shortest possible computational time (within an existing reconstruction framework). Reconstruction speed is also an important quality metric of how useful an image reconstruction algorithm is for clinical applications. Of course, in addition to algorithm design, software implementation of projectors and numerical operations have large impact on reconstruction speed. This would be an interesting aspect for future challenges, but would need careful design of the evaluation criteria as well as software infrastructure.

Use of numerical simulations instead of real data would have provided ground truth solutions. Nevertheless, we focused on the assessment of the performance of the algorithms on real data which makes the translational step from scientific development to application on clinically relevant data smaller. 

The focus of this challenge was on classical optimisation schemes to solve a well-defined optimisation problem. This allowed us to evaluate the different submissions even without having a ground truth solution. Nevertheless, this made this challenge less suited for deep-learning-based image reconstruction algorithms. While implementing learned reconstruction methods in the STIR, SIRF and CIL software is already feasible \cite{singh20233DPETDIPReconstruction}, current software limitations affect run-time. An important aim of our community is to extend STIR, SIRF and CIL to support memory-efficient integration with Python CUDA libraries such as PyTorch, before organising a challenge which is also suitable for deep-learning-based submissions. Such a challenge would likely concentrate on metrics evaluating image quality (as opposed to convergence), including lesion detectability, contrast and noise. Due to the nature of deep-learning methods, data from most ``standard" phantoms is not suitable, and will likely have to be replaced by a combination of simulated and measured clinical data, with associated difficulties in data access.

Given the successes of PETRIC, we have launched \href{https://github.com/SyneRBI/PETRIC2/wiki}{PETRIC2}, which is open at the time of writing and is similar to PETRIC but uses data at reduced count level. Additional challenges are also considered, for example by adapting the methodology for SPECT reconstruction. 

\section{Conclusion}
In this article, we presented PETRIC, the first challenge for PET image reconstruction using measured PET datasets from a range of PET scanners. Its solid foundations allow further adaptation for different types of evaluation of the performance of image reconstruction methods on existing or new datasets from any PET scanners, radionuclides, and phantoms. 

\section*{Conflict of Interest Statement}

The authors declare that the research was conducted in the absence of any commercial or financial relationships that could be construed as a potential conflict of interest. CT has received funding for other research projects from General Electric (GE) HealthCare, Siemens Healthineers and Positrigo. He also holds patents with GE HealthCare and a scientific advisory role with Positrigo. KT has received funding for other research projects from General Electric (GE) HealthCare, Siemens Healthineers, and has a scientific advisory role with Positrigo.

\section*{Author Contributions}
All authors contributed in the article content and reviewed its components with variable contributions in each section. Also, all authors were involved in the design and/or implementation of the PETRIC challenge.

\section*{Funding}
PETRIC was organised by members of the Collaborative Computational Project in Synergistic Reconstruction for Biomedical Imaging (CCP SyneRBI) which has been funded by grants of the Engineering and Physical Sciences Research Council (EP/T026693/1) and by the Science and Technology Facilities Council through the UK Research and Innovation Digital Research Infrastructure program. CT is funded by the NWO AES Vici Talent program (doi:10.61686/REEAR02777). 

\section*{Acknowledgments}
We are grateful to all data contributors whose contributions have been essential for the training and completion of the challenge, including Jorge Cabello (Siemens Healthineers), Daniel Deidda (NPL), Kuan-Hao Su (GE Healthcare), Markus Jehl (Positrigo), Nicole Jurjew (UCL), Francesca Leek (UCL), Zekai Li, Philipp Mohr, Laura Providencia (UMCG), Pawel Markiewicz (LSBU), and Jarmo Teuho (Turku PET Centre). We wish to thank all parties that have participated in PETRIC as well as the organisers of the 2024 IEEE Nuclear Science Symposium and Medical Imaging Conference and the speakers at the PETRIC workshop. Finally, PETRIC would not have been possible without the many contributors to the open source software projects used for this challenge.


\section*{Data Availability Statement}

The curated and processed datasets used for PETRIC are listed in Table \ref{tab:training_data} and \ref{tab:evaluation_data}. 
These curated datasets are available  at \href{https://petric.tomography.stfc.ac.uk/data}{petric.tomography.stfc.ac.uk/data}, and processing scripts can be found in the SyneRBI/PETRIC GitHub repository \href{https://github.com/SyneRBI/PETRIC/tree/main/SIRF_data_preparation}{here}.



\bibliographystyle{Frontiers-Vancouver} 
\bibliography{references}

\section*{Appendix}
This appendix gives example steps followed during the data curation/preparation process. All data are normally placed in \verb|~/devel/PETRIC/data/<datasetname>| with datasetname following the convention of \verb|scannername_phantomname|.

If starting from Siemens mMR list-mode data, SIRF takes care of scatter estimation and other calculations.  An example script for preparation of the Siemens mMR NEMA IQ
dataset is \href{https://github.com/SyneRBI/PETRIC/blob/main/SIRF\_data\_preparation/Siemens\_mMR\_NEMA\_IQ/prepare\_mMR\_NEMA\_IQ\_data.py}{prepare\_mMR\_NEMA\_IQ\_data.py}. Output
of this script includes the projection data $y$ (\texttt{prompts.hs}), multiplicative data $m$ (\texttt{mult\_factors.hs}), additive term $a$ (\texttt{additive\_term.hs}), see eq.~\ref{eq:forward_model}. For some datasets, e.g. Siemens mMR ACR data, the preparation steps followed a modified workflow to align the available attenuation image to the SIRF OSEM image. In initial step, data were reconstructed without attenuation and scatter correction were not performed during this stage. This was used to register the mu-map to the reconstructed image. The output included aligned mu-map files stored in the processing directory. The reconstruction process was then repeated, now including attenuation estimation and scatter correction. For other datasets, we relied on vendor-supplied software to generate the projection data and terms in the forward model.

Steps below create the reconstructed OSEM image (\texttt{OSEM\_image.hv}), the penalty voxel-dependent weights (\texttt{kappa.hv}, eq. \ref{eq:RDP}) and $\beta$ (\texttt{penalisation\_factor.txt}, eq.~\ref{eq:objective_function}) and various plots.

\begin{itemize}
\item Change working directory to where data reside and add PETRIC to your python-path, e.g.,
\begin{verbatim}
PYTHONPATH=~/devel/PETRIC:$PYTHONPATH
\end{verbatim}
\item Run initial \verb|data_QC.py|:
\begin{verbatim}
python -m SIRF_data_preparation.data_QC    
\end{verbatim}
\item Run \verb|create_initial_images.py|:
\begin{verbatim}
python -m SIRF_data_preparation.create_initial_images \
    --template_image=<some_image>
\end{verbatim}
where the template image is one of the given VOIs (it does not matter which one, as they should all have the same size). (If you need to create VOIs yourself, you can use \verb|None| or the vendor image).
\item Edit \verb|OSEM_image.hv| to add modality, radionuclide and duration info which got lost (copy from \verb|prompts.hs|)
\item Edit \verb|dataset_settings.py| for subsets (used by our reference reconstructions only, not by participants).
\item Edit \verb|petric.py| for slices to use for creating figures (\verb|DATA_SLICES|). Note that \verb|data_QC| outputs centre-of-mass of the VOIs, which can be helpful for this.
\item Optionally make VOIs, e.g.,
\begin{verbatim}
python -m SIRF_data_preparation.create_Hoffman_VOIs \
    --dataset=<datasetname>
\end{verbatim}
\item Run \verb|data_QC.py| which should now make more plots. Check VOI alignment etc.
\begin{verbatim}
python -m SIRF_data_preparation.data_QC --dataset=<datasetname>
\end{verbatim}
\item Get penalisation factor by comparing to a dataset from the same scanner, e.g.
\begin{verbatim}
python -m SIRF_data_preparation.get_penalisation_factor \
    --dataset=NeuroLF_Esser_Dataset \
    --ref_dataset=NeuroLF_Hoffman_Dataset -w
python -m SIRF_data_preparation.run_OSEM <datasetname>
\end{verbatim}
\item Run BSREM to generate reference solution. You probably want to monitor how these images look like as the recon will take a long time
\begin{verbatim}
python -m SIRF_data_preparation.run_BSREM  <datasetname>
\end{verbatim}
\item Adapt \verb|plot_BSREM_metrics.py| (probably only the \verb|<datasetname>|) and run interactively.
\item Copy the BSREM \verb|iter_final| to \verb|data/<datasetname>/PETRIC/reference_image|, e.g.,
\begin{verbatim}
stir_math data/<datasetname>/PETRIC/reference_image.hv \
    output/<datasetname>/iter_final.hv
\end{verbatim}
\item Clean-up some files
\begin{verbatim}
cd data/<datasetname>
rm -rf *ahv PETRIC/*ahv output info.txt warnings.txt
\end{verbatim}
\item Perform final check, such as its \verb|README.md| etc.
\item Transfer to web-server.
\end{itemize}

\end{document}